Submitted to Phys. Rev. E. Postscript figures are tarred, compressed  
and uuencoded. They appear after a ----cut here---- line.

\documentstyle[pre,aps,preprint,eqsecnum,epsf]{revtex}

\begin{document}
\draft
\preprint{patt-sol/9507004}

\title{\bf Front Stability in Mean Field Models of Diffusion Limited  
Growth}

\author{Douglas Ridgway  and Herbert Levine}
\address{Department of Physics and Institute for Nonlinear Science,  
University
of California, San Diego, La Jolla CA 92093-0402}
\author{Yuhai Tu}
\address{IBM T.~J.~Watson Research Center, PO Box 218, Yorktown  
Heights, NY
10598}
\date{\today}
\maketitle

\begin{abstract}
We present calculations of the stability of planar fronts
in two mean field models of diffusion limited growth.
The steady state solution for the front can exist for
a continuous family of velocities, we show that the selected
velocity
is given by marginal stability theory. We find that naive mean field  
theory has
no instability to transverse perturbations,
while a threshold mean field theory has such a Mullins-Sekerka  
instability.
These results place on firm theoretical ground the observed lack of  
the
dendritic morphology in naive mean field theory and its presence in  
threshold
models.
The existence of a Mullins-Sekerka instability is related to the  
behavior
of the mean field theories in the zero-undercooling limit.
\end{abstract}

\narrowtext

\section{Introduction}
Some of the most appealing and difficult
patterns found in nature are formed by diffusion limited growth.
A canonical example is solidification patterns  
\cite{KKL,solidification},
which, in addition to generating the delicate beauty of snowflakes,   
are
important metallurgically as the source of alloy microstructure.   
Diffusion
limited growth drives pattern formation in a wide variety of systems:   
examples
include fingering patterns
in  the displacement of a viscous fluid \cite{KKL},
 electrodeposition patterns  
\cite{electrodeposition,electrodeposition2},
microbial aggregation \cite{bacteria,bacteria2},  dielectric  
breakdown,  flow
in oil reservoirs, and possibly the wiring of neuronal dendritic  
arbors
\cite{HS}.

In recent years, much progress has been made on the problem of  
constructing a
theory of pattern formation in dissipative systems \cite{Cross}.   
This theory
explains the universal patterns seen in systems where a linear  
instability of a
uniform ground
state saturates at some finite amplitude.
In this context,  diffusion limited growth patterns form an  
interesting example
of a pattern forming system whose instability never
saturates.\footnote{Although situations can be created where some of  
the theory
of weakly nonlinear pattern formation can be applied,  such as  
directional
solidification\cite{KKL}.}

The instability which drives pattern formation in diffusion limited  
growth is
the Mullins-Sekerka instability.
The physical nature of this instability can be seen by
considering a solid seed particle immersed in a supercooled liquid  
bath. As
the liquid solidifies around the seed, the latent heat released by  
the first
order phase transition must diffuse away.
An outward perturbation on the boundary of the growing solid can cool  
more
easily, and so grows faster than neighboring regions, leading to an  
instability
of the solid-liquid interface.
Competition between the Mullins-Sekerka instability and other  
effects,  such as
surface tension,  sets a basic length scale for pattern formation.
As boundary features formed by the Mullins-Sekerka instability  
eventually grow
to a size where they too are unstable, new structures grow on them,  
and highly
complex fractal patterns are formed.

Diffusion limited growth occurs in at least two broad morphologies:  
tip
splitting growth, as seen for example in Hele-Shaw flow in a radial  
geometry,
and dendritic growth, as observed in solidification.
In a radial geometry, tip splitting growth forms branching fingers  
whose
ensemble averaged envelope is circular,  while dendritic growth forms  
a concave
envelope lead by cusp-like tips propagating at some fixed velocity.
The mechanism acting at the tips to select the actual behavior is  
understood:
anisotropy in the surface tension (or phase transformation kinetics)  
acts as a
singular perturbation at the tip to stabilize the dendrite, with the  
unique
velocity being selected by a microscopic solvability condition  
\cite{KKL}.  If
anisotropy is absent,
no propagating dendrite can be formed, and tip splitting growth is  
observed.
\footnote{Isotropic growth fingers can also be stabilized by growth  
in a
sufficiently narrow channel \cite{KKL}.}

More recent work is on global aspects of these pattern forming
systems. There is numerical evidence
\cite{ShochetThesis,ShochetMorph,ShochetMorph2} for a morphology
transition between the concave asymptotic profiles of the dendritic  
growth
morphology and the convex asymptotic profiles of the densely branched
morphology (DBM).  The work of Shochet et al.~suggests that there is  
a
discontinuity in the slope of the velocity
vs.~control parameter curve at the convex to concave transition,  
implying
that the different morphologies are akin to thermodynamically  
different
states. This kinetic phase transition hypothesis is bolstered by  
their
suggestion
that there may be coexistence between the branching and dendritic  
morphologies
\cite{coexist}.

One step towards theoretical understanding of this morphology
transition and of
diffusion growth patterns in general is to formulate an envelope  
model
which displays the convex/concave transition of the underlying  
microscopic
models.  Such a model, based on the classic mean field models of  
Witten and
Sander \cite{WittenSander}, has been proposed  
\cite{BLT,BLT2,BLT3,Tu}.
Since a model for the envelope cannot allow concavity (or the  
dendritic
morphology) if it forms stable planar fronts (by the kinetic Wulff  
construction
discussed in Section~\ref{KineticWulff}), it is necessary
to look for a Mullins-Sekerka
instability in the models discussed in \cite{Tu}, and to understand  
the nature
of that instability.

In the remainder of this paper we will discuss several microscopic
solidification models in general and review the behaviors of the
mean field models (Section~\ref{MeanField}).
We will present some background on the geometrical construction of  
asymptotic
shapes (Section~\ref{KineticWulff}),  the relationship between  
morphology
transitions and  the Mullins-Sekerka instability, and the velocity  
selection of
planar fronts (Section~\ref{MSsection}).
Our calculation of the stability of planar fronts in mean field  
models will be
presented in Section~\ref{MSstab}. The zero undercooling limit will  
be
discussed in Section~\ref{Zero},
and our conclusions appear in Section~\ref{lastSec}.

\section{Models and Mean Field Theory}
\label{MeanField}
Modelling efforts for diffusion limited growth have taken a variety  
of forms.
The direct  continuum approach for a sharp interface is in terms of a  
moving
boundary value problem \cite{KKL}.  The field controlling growth  
($u$, which is
proportional to the undercooling $T_m - T$) satisfies the diffusion  
equation
\begin{equation}
\label{continuum1}
\partial_t u = D \nabla^2 u \mbox{  .}
\end{equation}
The boundary conditions are undercooling at infinity and a melting  
point on the
boundary set by a surface tension $\gamma$ (due to the Gibbs-Thompson
condition) and phase transformation kinetics
\begin{equation}
\label{continuum2}
\begin{array}{rcl}
u|_{+\infty}  & = & \Delta \\
u|_{\mbox{interface}} & = & - \gamma \kappa - \beta v_n \mbox{  .}
\end{array}
\end{equation}
The boundary motion  is determined by the local gradient of the  
diffusing field
\begin{equation}
\label{continuum3}
v_n = \nabla u |_{\mbox{interface}} \cdot \hat{n}
\end{equation}
where $\hat{n}$ is the normal unit vector of the interface.
The displacement of a more viscous fluid by a less viscous fluid (the
Saffman-Taylor problem) or flow in porous media satisfies similar  
equations,
with the field $u$ representing pressure in the fluid\cite{KKL}.

More flexibility is offered by  phase field models, which  specify a  
continuous
transition from the solid to liquid phase
\cite{PhaseField,PhaseField2,PhaseField3}, allowing arbitrary  
topology of the
interface and more control over
the boundary layer and the phase transformation
at the expense of an additional field.

Another approach is the diffusion limited aggregation (DLA) class of  
models
\cite{WittenSander,Uwaha}, which have generated a great deal of  
interest.
This approach models diffusion limited growth by the discrete  
aggregation
of random walkers,  with the probability density of a random walk
capturing diffusive behavior in a computationally efficient scheme.
The random fractal nature of these aggregates motivate questions  
about
the effects of noise.
A phase field model with stochastic properties is studied in  
\cite{Keblinski}.
Other models include the cell dynamical scheme \cite{LiuGoldenfeld}  
and
the diffusion transition scheme \cite{ShochetMorph},  both of which  
couple
numerical solution of the diffusive field with some model for the  
phase
transition in cells on a lattice. Noise is an inherent part of these  
models as
well.

The various different growth morphologies may all be
observed within any particular model,  depending on
such parameters as boundary conditions,  undercooling, surface  
tension,
kinetic effect, anisotropy, etc.
The dendritic morphology occurs at finite
anisotropy, finite undercooling; while single walker DLA corresponds  
to
zero undercooling, zero surface tension and kinetic effect. Tip  
splitting
growth as seen in
radial Hele-Shaw flow and
Saffman-Taylor fingers in a channel geometry occurs at zero  
anisotropy.

\subsection{Mean field theory}
Mean field models attempt to capture global morphology of diffusion  
limited
growth.
Naive mean field theory in open radial geometries at zero  
undercooling has a
spherically symmetric solution which captures the mass scaling  
behavior of
single walker DLA, giving a mean field exponent of $d_f = d - 1$, a  
number
believed to become exact as $d \rightarrow \infty$.  However, this
spherically symmetric solution is always unstable, leading to  
difficulties
of interpretation. In addition, the naive mean field theory fails  in  
one
dimension or in a channel geometry,  where fronts accelerate to  
infinite
velocity in finite time \cite{BLT}.  A phenomenological modification  
of
the naive MFT has been proposed which fixes these problems and  
successfully
explains several important morphological observations.  One is that  
the
envelope of ensemble averages of DLA in a channel geometry takes the  
shape
of a Saffman-Taylor finger \cite{BLT}, and another is that  
anisotropic
growth in a radial geometry can form the global dendritic
morphology \cite{Tu}.
There are now indications that the necessary threshold suggested
phenomenologically in \cite{BLT}  can be generated by including the  
essential
effects of noise in underlying model \cite{TL}.

We begin our discussion of mean field theory by considering the
Witten-Sander aggregation model with finite walker density.
In this model,  the diffusion field controlling growth is modelled by  
random
walkers.
When a walker becomes adjacent to the solid,  it transforms into  
solid and
advances the solid to  that position.
This sticking rule implies that the walker density $u$ on the  
boundary is zero,
giving a walker density adjacent to the boundary of $a \nabla u$.  
This gives a
growth velocity proportional to $\nabla u$.  Comparison with  
(\ref{continuum2})
and (\ref{continuum3}) shows that DLA is a stochastic model for  
diffusion
limited interface growth with no surface tension or interface  
kinetics.
 The random walker density will satisfy the
diffusion equation, minus the walkers lost to the growing aggregate  
density
$\rho$
\begin{equation}
\dot  u  =  D \nabla^2 u - \dot \rho \mbox{ .}
\end{equation}
The transition probability of an unoccupied site ${\bf x}$ is  
proportional to
\begin{equation}
 u{\bf (x},t) P({\bf x}, t)
\end{equation}
where $P({\bf x})$ represent the probability of the neighboring sites
to be occupied by the aggregate.
We can write $P({\bf x})$ in the form
\begin{equation}
P({\bf x}) =  \sum_{i}\rho({\bf x}+{\bf e}_i)
\end{equation}
where ${\bf e}_i$ are lattice vectors to adjacent sites and $i$ runs  
over the
number of nearest neighbors. This formulation allows multiple walkers  
to occupy
a single site,  a feature
not  expected to affect the universality class.

We derive a continuum mean field model by allowing $u$ and $\rho$ to  
take on
any
continuum values and identifying the sum of nearest neighbor terms in  
$P$ as a
lattice Laplacian.
This  results in the Witten-Sander mean field model for finite   
walker density
DLA:
\begin{eqnarray}
\label{NMFT1}
      \dot  u & = & D \nabla^2 u - \dot \rho  \\  \label{NMFT2}
    \dot \rho & = & K u(\rho + a^2 \nabla^2 \rho) \mbox{  .}
\end{eqnarray}
where the constant parameters $D$, $K$ and $a$ can be removed by  
setting $ x
\rightarrow {ax}$, $t \rightarrow t a^2/D$,  and $\{\rho, u\}  
\rightarrow
\{\rho, u\} D/a^2 K$.
In these units the equations become
\begin{eqnarray}
\label{WSnonzero}
  \dot u  & = &  \nabla^2 u - \dot \rho \\
    \dot \rho & = &  u(\rho +  \nabla^2 \rho)
\end{eqnarray}
with a boundary condition on the undercooling, or walker density  at  
infinity
$u|_{+\infty} =\Delta $.  The total number of particles (aggregated  
and
walking)  $\rho + u$ is conserved by the local dynamics.

This mean field theory has no threshold for aggregate growth,  so a  
small
perturbation in the $\rho$ field, in the presence of $u$, will grow
exponentially. However, lattice simulations with discrete phase  
states,
including the DLA model discussed above as well as others
\cite{LiuGoldenfeld,WittenSander,ShochetMorph},  have an implicit  
threshold for
growth:
growth at a site is disallowed unless a nearest neighbor site is  
fully
occupied.  Phase field models also have a threshold, in this case  
explicit in
the structure of the form of the free energy,  which specifies that  
the phases
are at least metastable \cite{Keblinski,PhaseField}.
This lack of a threshold in naive MFT leads to problems in
channel or planar geometries, namely the infinite propagation  
mentioned
earlier.
Inserting a threshold by hand, we have the
phenomenological model introduced by Brener et al. ~\cite{BLT}:
\begin{equation}
\label{BLTmodel1}
\begin{array}{rcl}
      \dot  u & = & \nabla^2 u - \dot \rho  \\
    \dot \rho & = & u(\rho^\gamma + \nabla^2 \rho) \mbox{ .}
\end{array}
\end{equation}
Any function $F(\rho)$ which vanishes faster than linearly as $\rho  
\rightarrow
0$ could be used in the place of $\rho^\gamma$, for example
$F(\rho) = \rho \Theta(\rho- A)$.

The success of the $\gamma$ model motivates  the question of how to  
obtain such
a growth threshold in a more fundamental way.  One possibility is to  
invoke the
generic effects of noise \cite{TL}.  In microscopic theory,  thermal  
noise at
the tip drives sidebranching \cite{KKL}.  In the Witten-Sander  
particle
aggregation model,  shot noise of particle diffusion and aggregation  
are a
source of noise.  Noise is known to shift the bifurcation point of  
nonlinear
systems \cite{Graham,Shapiro}.

To sketch how noise may affect the form of the macroscopic mean field
equations, recall the derivation of the naive mean field theory  
(\ref{NMFT1},
\ref{NMFT2}) from the Witten-Sander rules.  The transitions of  
particles on a
nearest neighbor site from walker to aggregate are random and  
independent,  and
therefore governed by  Poissonian shot noise.  Including these  
fluctuations by
a Langevin noise source $\eta$ multiplicatively coupled with the  
field, we have
\begin{equation}
    \dot \rho  =  K u(\rho + a^2 \nabla^2 \rho)  + \sqrt{ K u(\rho +  
a^2
\nabla^2 \rho)}\  \eta
\end{equation}
Multiplicatively coupled noise can lead to stabilization of an  
unstable system,
 analogous in some ways to a parametrically driven pendulum  
\cite{Shapiro}.
This stabilization is exactly what is required,  as it establishes a  
lower
threshold in $\rho$ below which  growth does not occur.  A more  
complete
discussion appears in \cite{TL}.

\section{Theory of kinetic morphology}
\label{KineticWulff}
In this section, we review some material concerning the growth of  
envelopes.
A system with a transition layer that does not spread will in the  
long time
limit be described
by a sharp interface model. We parameterize the interface between the  
two
phases
by a curve $\vec x(\sigma, t)$
and write the evolution of this curve in the form
\begin{equation}
\label{dxdt}
{ d \vec x\over dt} = \vec v(\sigma, t, [\vec x])
\end{equation}
where $\vec v$ can in general be a complicated functional depending  
nonlocally
on the entire curve
$\vec x$.
We are interested in envelopes which form a well defined shape in the
asymptotic time limit,  that is assume a scaling form
\begin{equation}
\label{shapeScale}
\vec x(\sigma, t) = R(t) \vec r(\sigma) \mbox{ .}
\end{equation}
Since velocity parallel to the interface corresponds merely to a  
redefinition
of the parameterization $\sigma$, we take $\vec v = v_n \hat n$. From
(\ref{dxdt}) and (\ref{shapeScale}) we get
\begin{equation}
{ d \vec x\over dt} = \dot R(t) \vec r(\sigma) = v_n(\sigma, t) \hat  
n
\end{equation}
If the envelope scales, the time dependence will factor out of $v_n$,  
leaving
us with
\begin{equation}
\label{shapeScale2}
v_n(\sigma) = \vec r(\sigma) \cdot \hat n \mbox{ .}
\end{equation}
One may imagine several possibilities for the nature of $v_n$.
For example, in a sharp interface model for diffusion limited growth,
$v_n$ will depend on the whole structure and history of the envelope  
in some
nonlocal way.
Alternatively, it may be that $v$ depends only on purely local  
properties of
the interface, such as the normal direction, curvature and so on. In  
the
asymptotic limit, the envelope locally appears to be flat, leaving  
the normal
direction as the only remaining
variable that $v$ can depend on. In this case, one can solve for the  
envelope
shape explicitly
given $v_n(\hat n)$ for all angles of flat front propagation. This is  
done by
the kinetic Wulff construction to be discussed next.

The equilibrium shape of a crystal has a solution in the form of the  
Wulff
construction which dates back to the turn of the century
\cite{Wulff,Dobrushin}.  Given the anisotropic form of the surface  
energy
$\gamma(\theta)$ for each crystal direction $\theta$, minimizing the  
total
surface energy (at fixed area) leads to the following construction:   
for every
normal direction,  draw a line perpendicular to the normal at a  
distance from
the origin proportional to surface energy of that direction.  The  
inner convex
hull of these lines is the equilibrium crystal shape.  The resulting  
crystal
may be faceted,  smooth (if it is above the roughening transition) or  
may have
both facets and smoothly curved surfaces.

If a crystal is grown under such conditions that the normal velocity  
depends
only on the direction of the surface normal (for example,  in a  
reaction
limited regime),  then the steady state shape is again given by the  
Wulff
construction,  with normal velocity substituted in for surface  
energy.  It is
clear that such a  steady state cannot be concave  
\cite{Wolf,Wettlaufer}.
Because the envelope of the dendritic morphology is concave
\cite{ShochetMorph2},  a geometrical model cannot capture the  
dendritic
morphology.

Although the kinetic Wulff construction is not applicable to the  
nonlocal
diffusion limited regime,  it is important to understand its  
consequences for
general modelling of the morphology transition.  A model which forms  
stable
planar fronts at any angle in two dimensions cannot capture the  
dendritic
morphology.
In the asymptotic limit, a scaling form will arise, composed of  
approximately
planar fronts.
The velocities of these planar fronts as a function of angle can be  
found by
marginal stability.
The asymptotic shape will therefore be given by the kinetic Wulff  
construction,
and will be
convex.
So a mean field theory which always has a stable planar front  cannot  
model the
dendritic (concave) morphology, and cannot describe the morphology  
transition.
In fact, in the convex regime, the planar front has to have an  
instability
analogous to Mullins-Sekerka instabilty.
It is for this reason that we have tested for the presence of a  
Mullins-Sekerka
instability in mean field models.

In our calculations we have not  included a term such as
\begin{equation}
\epsilon \left[ \frac{\partial^4 \rho}{\partial x^4} +  
\frac{\partial^4
\rho}{\partial y^4} \right]
\end{equation}
to explicitly include anisotropy.
Although such an anisotropy term is of course a necessary ingredient  
for
dendritic growth,
we have neglected it for simplicity.
Its effect will be small; its function at the level of our linear  
stability
analysis would merely be to pick out a preferred direction for the  
instability
to grow.
This point of view is verified in 2d simulations, which show  
explicitly the
existence of dendritic growth in exactly those systems with a  
Mullins-Sekerka
instability in isotropic planar fronts\cite{Tu}.

\subsection{Marginal stability}
\label{MSsection}
To find the front velocity $v(\theta)$ used above, it is necessary to  
find the
velocities of  planar fronts in the desired model.
The mean field diffusion growth equations derived earlier
have the generic property (shared with other nonlinear diffusion  
equations)
that there exist steady state solutions of any velocity. This  
property is
connected with the unstable nature of the invaded state, so it does  
not
apply to models with a strict cutoff, although it does apply to the  
$\gamma$
model with its power law cutoff.
The first task is therefore the solution
of a velocity selection problem.
Marginal stability theory states that, for typical initial  
conditions,
the selected
velocity is the lowest velocity for which localized perturbations are  
stable
in the moving reference frame\cite{MarginalStability,Cross}.
Here, ``localized'' means perturbations whose asymptotic tails decay  
no slower
than the front itself,  and are therefore smaller than the front  
everywhere.
Marginal stability is on a firm foundation for nonlinear diffusion  
problems
with one variable.
Marginal stability falls into
two classes: linear marginal stability, where computing the stability
of the asymptotic tail is sufficient, and nonlinear marginal  
stability, where
the (linear) stability of the entire front needs to be considered.
In one variable velocity selection problems, such as the Fisher  
equation $\dot
u = u(1-u) + \nabla^2 u$,  a parameter counting argument indicates a  
continuum
of unstable modes in the regime where the front
oscillates\cite{MarginalStability}, i.e.\ where $c<1/2$. In problems
involving nonlinear marginal stability, it is the appearance of a  
discrete
unstable mode which
sets the marginally stable velocity.

There is a qualitative distinction between fronts selected by linear  
and
nonlinear marginal stability. In either case the front is formed as a  
balance
between the growth of the density
and its diffusion which catalyzes further growth. In the linear case,  
the
growth of the front at
its leading edge is sufficient to catalyze its continued propagating,  
in this
case we speak of
the front being ``pulled'' by its leading edge. In the nonlinear  
case, the
linear growth rate
of the leading edge is insufficient to maintain the propagation  
velocity,
 and the larger nonlinear growth in the body of the front is  
necessary to
catalyze the continued propagation. In this case, we speak of the  
front being
``pushed'' by the growth behind the front \cite{pushPull}.

To compare with the one dimensional marginal stability theory we  
compute, for
the naive ($\gamma=1$) model,
the asymptotic behavior of the front at $z=+\infty$. This  can be  
found by
setting $u=\Delta$ in the equations for a steady state front (see
Section~\ref{Perturb}) giving
\begin{equation}
-c \rho' =  \Delta (\rho +  \rho'') \mbox{  .}
\end{equation}
This has solutions $\rho \sim \exp{\lambda_\pm z}$ where  
$\lambda_{\pm}$
is given by
\begin{equation}
\lambda_{\pm} = -{ c \over {2 \Delta}} \pm \sqrt{ { c^2 \over 4  
\Delta^2} - 1}
\mbox{  .}
\end{equation}
For $c<2\Delta$, these modes oscillate, which can be prohibited on
physical grounds by the interpretation of $\rho$ as a density.
The parameter counting argument analogous to the nonlinear diffusion  
case is
complicated by an additional variable and an additional free  
parameter
($\Delta$)  but the result is the same: namely, that we have a  
continuum of
instabilities in the regime where the fields acquire oscillating  
tails,
predicting a linear marginally stable velocity of $c=2\Delta$.

For  $\gamma>1$, we expect similar behavior --- the selected velocity  
will be
the slowest stable
velocity, which will again be the lowest velocity for which $\rho$  
does not
cross zero.
In this case, however, this will not be given by the behavior of the  
asymptotic
tail.  The linear
growth rate for $\gamma>1$ is zero,  and therefore any front formed  
must be
pushed, with a
velocity given by nonlinear marginal stability.

\section{Perturbation of moving fronts}
\label{Perturb}
We will write out the equations describing perturbations of fronts  
satisfying
(\ref{BLTmodel1}).
We consider a planar geometry, so that the only spatial dependence is  
$x$
dependence, and we will add transversal dependence back in later.
Working in a moving frame $z = x - c t$ and looking for a steady  
state solution
$\rho_t = u_t = 0$
we have
\begin{eqnarray}
- c u'  & = & u'' - c \rho' \\
\label{rho0eqn}
 - c \rho' & = & u(F(\rho) +  \rho'')
\end{eqnarray}
where primes indicate $z$ derivatives.
By integrating and using the boundary conditions  $\rho = \Delta$, $u  
= 0$ at
$z= -\infty$
the first equation becomes
\begin{equation}
\label{u0eqn}
 u' = c (\Delta - \rho - u)  \mbox{  .}
\end{equation}
By integrating forward from $z=-\infty$, a steady state front can be  
found for
any desired velocity.

The base solutions may be found by numerically integrating the  
equations.
Initial conditions at the left  are chosen to assure the
boundary conditions $u \rightarrow 0$ and $\rho \rightarrow \Delta$  
as $z
\rightarrow -\infty$. Since the $\rho$ field may become negative,
we use $\mbox{sgn}(\rho) |\rho|^\gamma$ as our threshold function.
 Next we do perturbation theory on our
equations in the moving frame
\begin{eqnarray}
 \dot u - c \partial_z u&& = \nabla^2 u - \dot \rho + c \partial_z  
\rho \\
\dot \rho - c \partial_z \rho &&= u(F(\rho) + \nabla^2 \rho)
\end{eqnarray}
where we include transversal dependence in our perturbation
\begin{eqnarray}
 u &&= u^0(z) + e^{\omega t} e^{i k y} u^1(z) \\
\rho &&= \rho^0(z) + e^{\omega t} e^{i k y} \rho^1(z)
\end{eqnarray}
and look for eigenfunctions $u^1$ and $\rho^1$ which go to zero as
$z \rightarrow \pm \infty $. Linearizing, we get
\begin{equation}
 \omega u^1 - c {u^1}' = -k^2 u^1 + {u^1}'' - \omega \rho^1 + c  
{\rho^1}'
\end{equation}
\begin{eqnarray}
 \omega \rho^1 - c {\rho^1}' &=& u^0(z)(F'(\rho^0(z)) \rho^1 - k^2  
\rho^1 +
{\rho^1}'')
 \nonumber\\
&& + u^1  (F(\rho^0(z)) + {\rho^0}''(z) )
\end{eqnarray}

Since there is little hope of solving this eigenvalue problem  
exactly,
we discretize it in a box and diagonalize it numerically.
To put this in a form which is easy to discretize, we write it as
\widetext
\begin{equation}
\label{perturb}
\left[ \begin{array}{cc}
      D^2 + c D - (F(\rho^0) + {\rho^0}'' + k^2) & -u^0 D^2 -
u^0 (F'(\rho^0)-k^2) \\
      (F(\rho^0) + {\rho^0}'' )   &  u^0 D^2 + c D  + u^0
(F'(\rho^0)-k^2) \end{array} \right]
\left[\begin{array}{c} u^1 \\ \rho^1 \end{array} \right]
= \omega \left[ \begin{array}{c} u^1 \\ \rho^1 \end{array} \right]
\end{equation}
\narrowtext
where $D$ represents the finite difference approximation of a  
derivative
expressed as a tridiagonal matrix.

Now, we also want our perturbation to decay as fast or faster than  
the
base solution at $z=+\infty$.
To do this, we can rewrite our eigenvalue problem in terms of new  
variables,
having divided out the asymptotic fall-off of the base solution tails  
on the
right, after which
a zero-slope boundary condition on the right enforces the
localized perturbation condition.
Explicitly, if we let $u^1_{old} = f_u(z) u^1_{new} $,
and $\rho^1_{old} = f_{\rho}(z) \rho^1_{new}$, where the $f$  
functions have
the same dependence at large $z$ as the base solution and approach a  
constant
at the left, then we have
\begin{equation}
D u^1 = D f_u(z) u^1_{new} = f_u(z) (\frac{f'_u}{f_u} + D) u^1_{new}
\end{equation}
and
\begin{eqnarray}
 D^2 u^1 &&=  f_u(z) (\frac{f'_u}{f_u} + D)^2 u^1_{new} \nonumber\\
      &&= f_u(z) (\frac{f''_u}{f} + 2 \frac{f_u'}{f} D + D^2)  
u^1_{new} \mbox{
.}
\end{eqnarray}
Making the replacements $D \rightarrow D +\frac{f'}{f}$ and $D^2  
\rightarrow
D^2 + 2\frac{f'}{f}D +\frac{f''}{f}$
in the perturbation equations (\ref{perturb}), we have
\widetext
\begin{eqnarray}
&&\left[ \begin{array}{c}
  D^2 + (c + 2 \frac{f_u'}{f_u})D + (-F(\rho^0)  -{\rho^0}'' -
                  k^2+ \frac{f_u''}{f_u} + c \frac{f_u'}{f_u} ) \\
  \frac{f_u}{f_{\rho}}(F(\rho^0) + {\rho^0}'' )  \end{array} \right.
\nonumber\\
  &&\left. \begin{array}{c}
        \frac{f_{\rho}}{f_u} \{-u^0  D^2  - 2  
\frac{f_{\rho}'}{f_{\rho}}
                 u^0 D - u^0 (F'(\rho^0) -k^2 +  
\frac{f_{\rho}''}{f_{\rho}})\}
\\
        u^0 D^2 + (c + 2 \frac{f_{\rho}'}{f_{\rho}} u^0) D  + u^0  
(F'(\rho^0)
-k^2 +
\frac{f_{\rho}''}{f_{\rho}} ) +
c \frac{f_{\rho}'}{f_{\rho}}
       \end{array} \right]
\left[\begin{array}{c} u^1 \\ \rho^1 \end{array} \right]
= \omega \left[ \begin{array}{c} u^1 \\ \rho^1 \end{array} \right]
\end{eqnarray}
\narrowtext
The functions $f$ are computed from the base solution for $\gamma >1$  
and
chosen analytically
for $\gamma=1$.
The growth rates of possible perturbations are just the eigenvalues  
of this
nonsymmetric matrix,  which can be found numerically by standard  
techniques.

\subsection{Numerical results}
\label{MSstab}

\begin{figure}[p]
\epsfxsize=3in
\epsffile{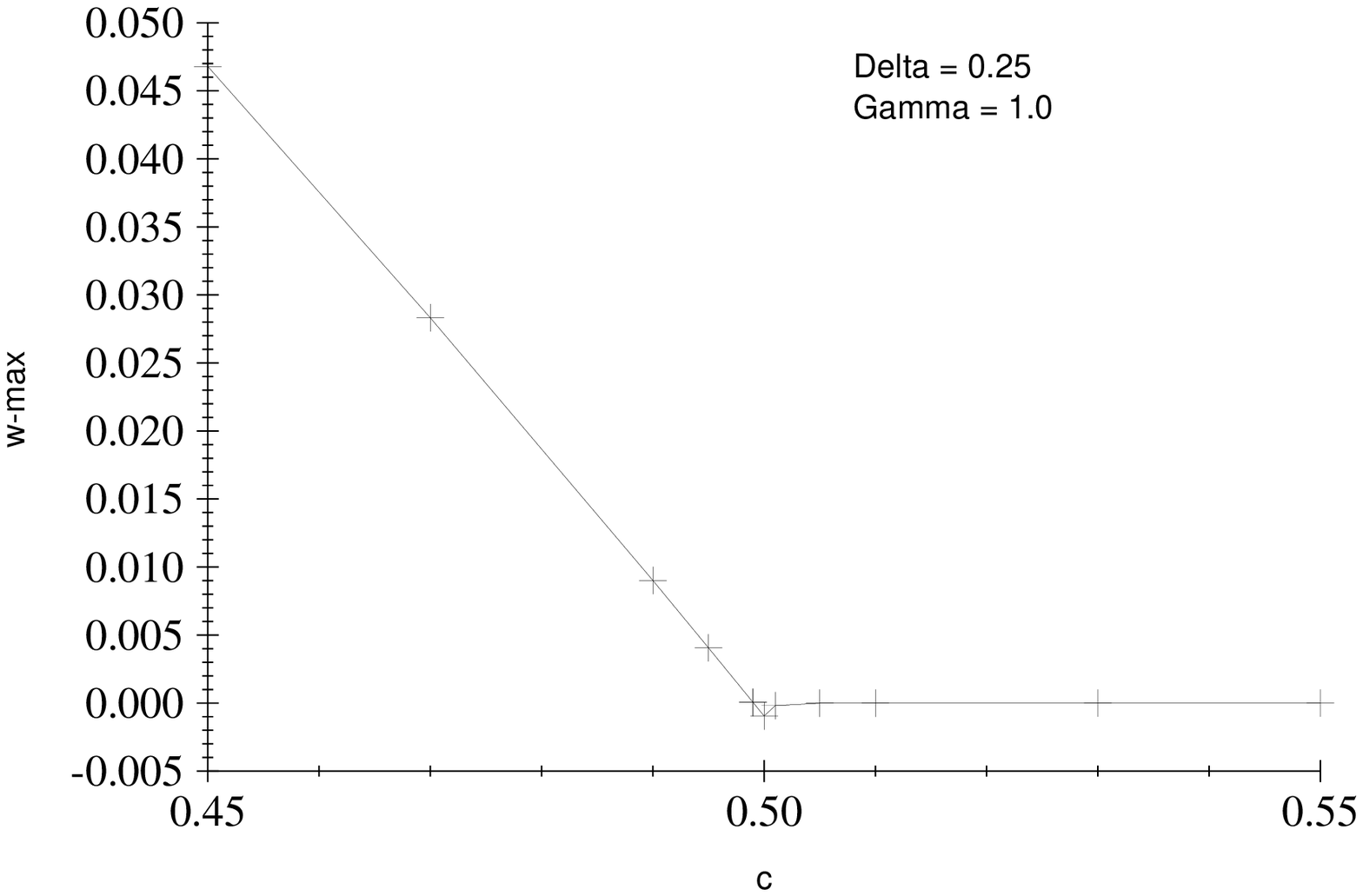}
\caption{Marginal stability plot for naive ($\gamma=1.0$) mean field  
theory
showing growth rate of most unstable mode vs. $c$. $\Delta = 0.25$.  
Below $c=2
\Delta = 0.50$, the $\rho$ field crosses zero and has an oscillating  
tail.The
displacement of the zero mode away from zero is due to the nature of  
the front
tails right at $c=2\Delta$, and does not occur for $\gamma>1$.}
\label{MSplot}
\end{figure}

We first display the results relating to velocity selection by
marginal stability.
The behavior of the largest growth rate $\omega_{\rm max}$ near  
$c_{\rm ms}=2
\Delta$  for $\gamma=1$ is shown in Figure~\ref{MSplot}. As expected,  
we see
that the growth rate for the most unstable mode drops to zero as $c$
passes through $2 \Delta$, implying a selected velocity of $2  
\Delta$. This
velocity selection
has been verified in direct numerical simulations.  At $\gamma=1$,  
then,
the front is pulled, and linear marginal stability holds.

\begin{figure}[p]
\epsfxsize=3in
\epsffile{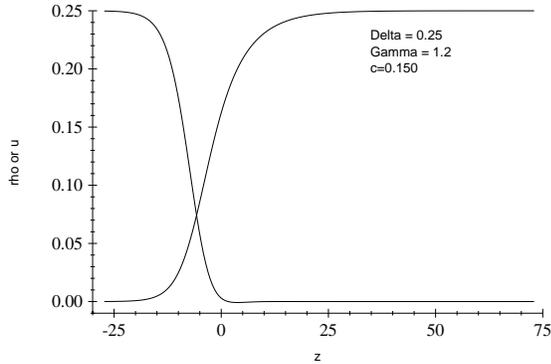}
\caption{$\rho$ and $u$ fields at a velocity below the marginally  
stable
velocity. Note how $\rho$ crosses zero. $\Delta = 0.25$,  
$\gamma=1.2$,
$c=0.15$.  At this $\gamma$ and $\Delta$, the selected velocity  
$c_{\rm ms} =
0.216$.}
\label{gFront}
\end{figure}

\begin{figure}[p]
\epsfxsize=3in
\epsffile{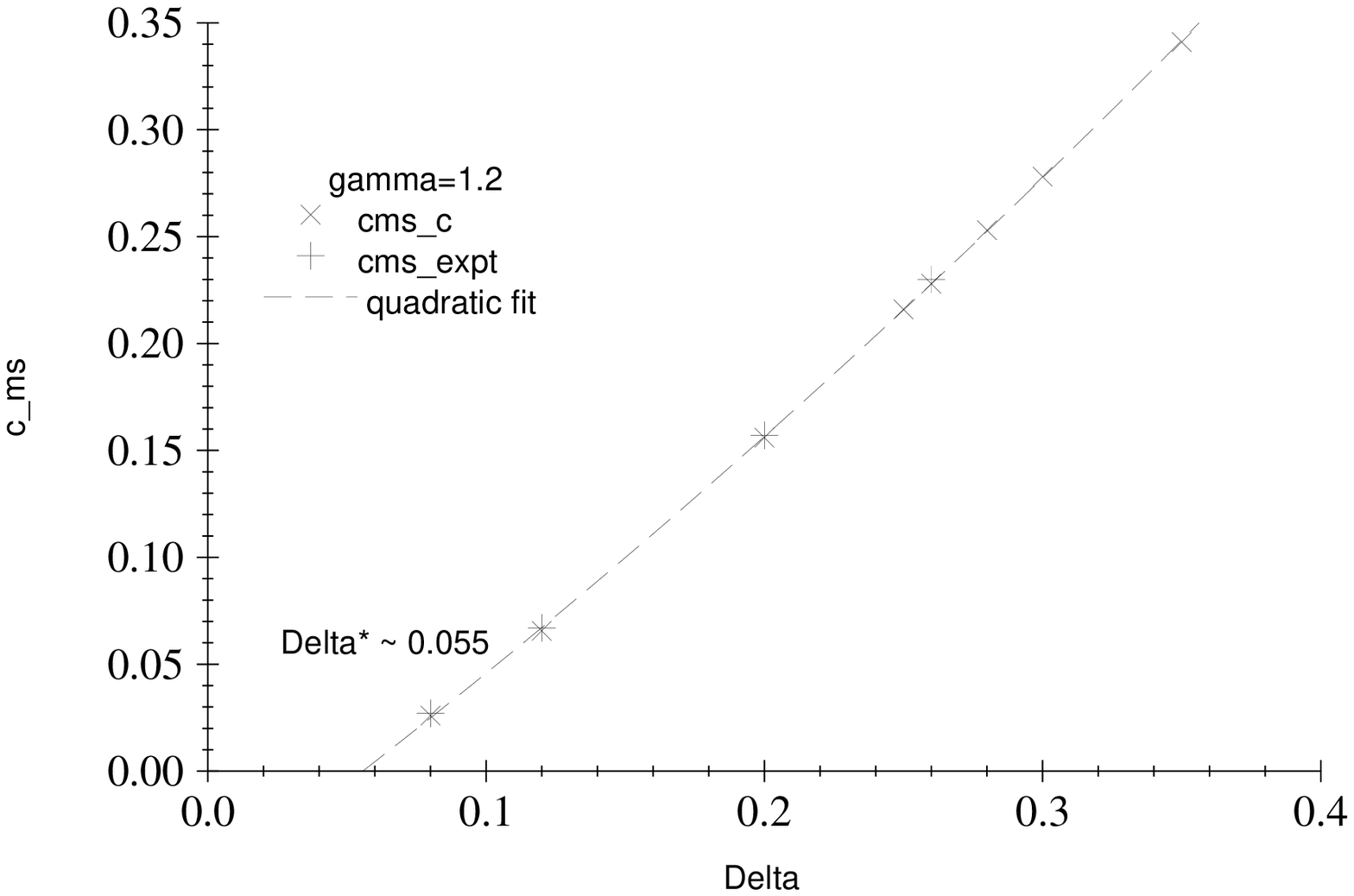}
\caption{Front velocity as a function of $\Delta$, as calculated by  
marginal
stability and measured by simulations. The marginally stable velocity  
is the
smallest velocity for which
the front is stable in its own frame against local perturbations.}
\label{VvsD}
\end{figure}

The profiles of the $u$ and $\rho$ fields at a velocity smaller than
$c_{\rm ms}$ for $\gamma>1$
are shown in Figure~\ref{gFront}.  At this velocity, the $\rho$ field  
crosses
zero to become negative. This creates a well allowing
a discrete unstable mode, analogous to a bound state in a well in  
quantum wave
mechanics.
For $\gamma>1$, we again expect the velocity to be given by the  
velocity of the
slowest stable
front.
This is now a global stability calculation for the velocity of a  
pushed front.
This velocity selection is verified in Figure~\ref{VvsD}, which  
compares
the velocity calculated from numerical diagonalization at  
$\gamma=1.2$ with the
velocity as measured in simulations.
It is clear that nonlinear marginal stability theory is predicting  
the correct
velocities as measured in simulations, and that the $\gamma$ model  
gives pushed
fronts.

\begin{figure}[p]
\epsfxsize=3in
\epsffile{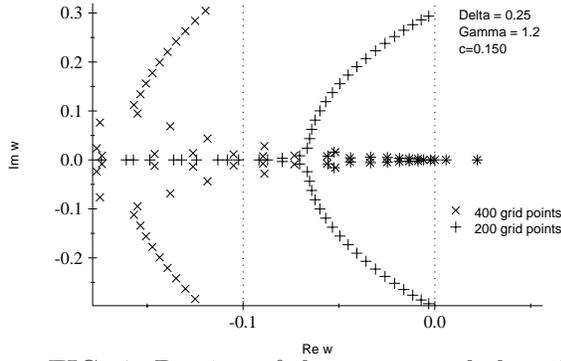}
\caption{Portion of the spectrum below the marginally stable  
velocity.
$\gamma=1.2$, $\Delta=0.25$. Two sizes of grids are shown, to give  
some
indication of the nature of the modes
in the continuum limit.  Most modes not shown lie on or close to the  
negative
real axis. }
\label{gSpec}
\end{figure}

The nature of the instability for $\gamma>1$ is indicated by the
spectrum shown
in Figure~\ref{gSpec}. As $c$ passes below $c_{\rm ms}$,
a single discrete mode
moves above zero. This is in contrast to the $\gamma=1$ case, where  
the top of
a continuum
moves above zero, but it is natural in light of the fact that $\rho$  
decays as
a power law and cannot have the oscillating tail that generates the  
continuum
in the $\gamma=1$ case.
Figure~\ref{VvsD} shows a quadratic extrapolation to zero velocity,  
which
occurs for $\gamma=1.2$ at $\Delta = \Delta^* \simeq 0.055$.  This  
defines
$\Delta^*$,
the lowest undercooling at which the model can form a propagating  
front.
 Nonzero $\Delta^*$ is the
essential feature of threshold mean field theories, and is not found  
in naive
mean field theory,
for which (since $c_{\rm ms} = 2 \Delta$) $\Delta^* = 0$.
The transverse stability calculation is done for the selected planar  
front, and
this
calculation gives the required front velocities.

\begin{figure}[p]
\epsfxsize=3in
\epsffile{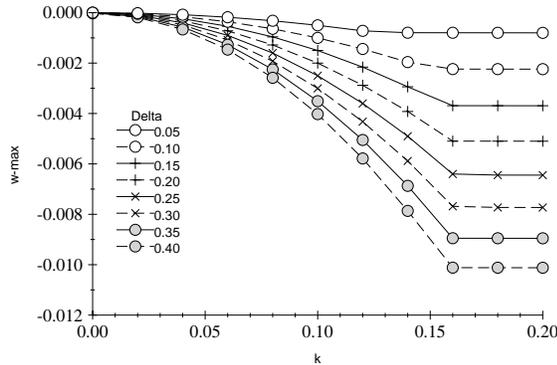}
\caption{Growth rate of most unstable mode as a function of  
transverse
wavenumber of perturbation,  Witten-Sander  
model~(\protect{\ref{WSnonzero}}).
The velocity is chosen to be $c = 1.02 c_{\rm ms} = 1.02 (2 \Delta)$.  
Similar
results are found
at $c=c_{\rm ms}$, except for the offset mentioned in
Figure~\protect{\ref{MSplot}}.}
\label{WSstabilityFig}
\end{figure}

Our main results concern the existence of a
Mullins-Sekerka instability \cite{MullinsSekerka} in these models.
Again we restrict ourselves to perturbations which decay at least as  
fast as
the front.
Figure \ref{WSstabilityFig} shows the growth rate $\omega$ of the  
most unstable
mode in naive
mean field
 model plotted as a function of the transverse wavenumber $k$ for a  
number of
undercoolings $\Delta$. There is no sign of any finite wavenumber  
instability.

\begin{figure}[p]
\epsfxsize=3in
\epsffile{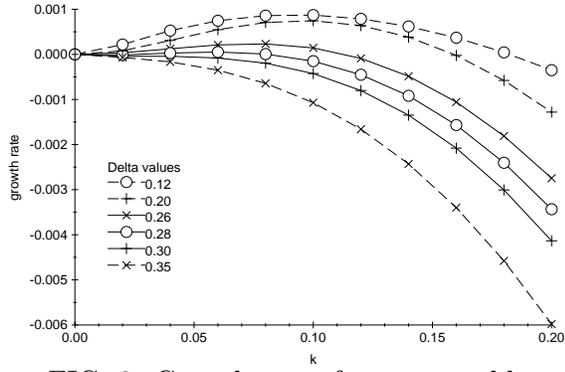}
\caption{Growth rate of most unstable mode as a function of  
transverse
wavenumber of perturbation,  $\gamma$  
model~(\protect\ref{BLTmodel1}),
 $\gamma = 1.2$ for various values of $\Delta$. Instability sets
in at $\Delta^{**} \approx 0.29$. }
\label{BLTstabilityFig}
\end{figure}

\begin{figure}[p]
\epsfxsize=3in
\epsffile{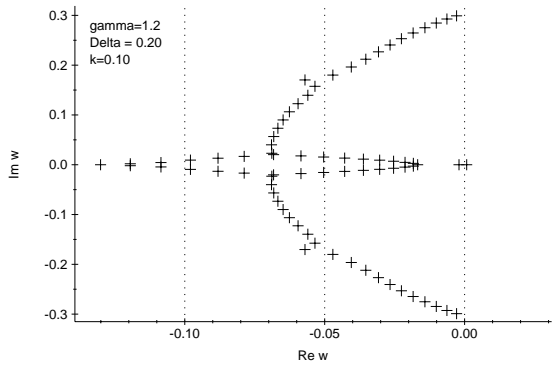}
\caption{A portion of the spectrum for $\gamma=1.2$, $\Delta = 0.20$,  
and $k =
0.10$.}
\label{kSpectrum}
\end{figure}

Figure \ref{BLTstabilityFig}  shows the largest growth rate $\omega$  
vs. $k$
for a number of $\Delta$'s in the $\gamma$ model at $\gamma=1.2$.
There is a finite $k$ instability  for $\Delta$ below $\Delta^{**}   
\simeq
0.29$. Figure \ref{kSpectrum} shows a typical unstable spectrum. The
characteristic length scale of the instability is
comparable to  the width of the front.
Comparing with Figure~\ref{VvsD}, note that there is no sign of any  
velocity
slope discontinuity at the
point $\Delta^{**} \simeq 0.29$ where the Mullins-Sekerka instability
disappears.
This argues against the morphology transition being some kind of  
nonequilibrium
phase transition.
These results demonstrate explicitly what was already indicated in  
\cite{Tu}: a
threshold is necessary to
generate a Mullins-Sekerka instability and allow concave dendritic  
envelopes to
be formed, and that the $\gamma$ threshold model in particular both  
has concave
envelopes and a Mullins-Sekerka instability.

\section{The limit of zero undercooling}
\label{Zero}
Undercooling is an important control parameter in diffusion limited  
growth in
general and the only control parameter in many walker DLA.
The diffusion transition scheme \cite{ShochetMorph}, for example,   
becomes
equivalent to the Ising model when the undercooling is maximal.
Many walker DLA is equivalent to the Eden model when the density of  
walkers
equals one.
The single walker DLA model of Witten and Sander \cite{WittenSander}
corresponds to the opposite limit of zero undercooling.
As undercooling is changed,  morphology transitions can be observed.  
Varying
the undercooling from high to low in the cell dynamical scheme,
Liu and Goldenfeld \cite{LiuGoldenfeld}
found transitions from dendritic growth aligned with crystalline axes  
to a more
or less isotropic state (interpreted as the dense branching  
morphology), and
then back to dendritic growth, this time with dendrites aligned  
between the
crystalline axes.

Naive mean field theory for single walker DLA takes the following  
form:
\begin{equation}
\label{WS0}
      \nabla^2 u = \dot \rho = u (\rho + \nabla^2 \rho) \mbox{ ,}
\end{equation}
where $u|_{+\infty} \sim J r$, i.e.\ there is some fixed flux of $u$  
at
infinity. This is the form in which DLA mean field theory first  
appeared \cite
{WittenSander},  however it is not a simple limit of the finite  
$\Delta$
theory.  To see the connection, we start with  
(\ref{NMFT1},\ref{NMFT2}) and put
$ u \rightarrow u/\Delta$,
$t \rightarrow \Delta t$, obtaining
\begin{eqnarray}
\dot \rho & = & u(\rho + \nabla^2 \rho) \\
\Delta \dot u  & = & \nabla^2 u - \dot \rho \mbox{  .}
\end{eqnarray}
The argument is now that as $\Delta \rightarrow 0$, we have $\nabla^2  
u = \dot
\rho$,  which  is the mean field model at zero undercooling quoted  
above.  The
question is what happens to the  boundary conditions.
We can relate the fixed flux boundary conditions of the usual zero  
undercooling
model
to the $\Delta \rightarrow 0 $ limit of the finite undercooling model  
by
considering the system in some box with a finite length $L$.
After the rescaling above, the boundary condition on $u$ is  
$u|_{+\infty}  =
1$.
In a system of length $L$, this corresponds to a supply of flux $J  
\sim 1/L$.
Taking the system size to infinity shows that  the direct $\Delta   
\rightarrow
0$ limit of the finite
undercooling model is the zero undercooling model with zero flux.
This somewhat odd condition is due to the nature of the Laplacian  
model:  it is
modelling infinitely slow growth,  so it must introduce a new  
artificial time
scale (given by the boundary flux $J$) on which growth formally  
proceeds.

 At first sight,
the lack of a Mullins-Sekerka instability in naive mean field theory  
seems
surprising,  as it seems to  contain all the ingredients which lead  
to a
Mullins-Sekerka instability in other diffusion limited growth  
systems.
In fact, there is a connection between the existence of a  
Mullins-Sekerka
instability and the
behavior of fronts in the zero undercooling limit alluded to earlier.
Specifically, any system with propagating fronts in the zero  
undercooling model
should,
in the finite undercooling version, 1) have a lowest undercooling  
$\Delta^*$,
below which propagating fronts cannot form and the boundary region  
spreads
forever and 2) have some region of instability above $\Delta^*$ and  
below some
value $\Delta^{**}$ where the system can form propagating fronts in  
one
dimension but where those fronts exhibit a transverse Mullins-Sekerka
instability.
These effects are both due to the existence of a threshold for growth  
in the
model.

For a zero undercooling model (such as the $\gamma$ model) which  
possesses a
travelling front solution, there is some selected value $\rho^o$ for  
the
density of the aggregate left behind by the front\cite{BLT}. This  
value is
independent of the boundary flux $J$, which can be scaled out by  
rescaling
time. Models such as the naive model do not have such an aggregate  
density; as
shown in \cite{BLT} via a scaling solution the frozen cluster scales  
with
distance as $1/x^2$, corresponding to a ``fractal dimension'' of   
$d_f = -1$.
The constant flux of walkers into the system forces the front to  
advance at an
ever-increasing rate, reaching infinite velocity in finite time.

Consider now a finite undercooling model as we approach the limit of  
zero
undercooling.  The aggregate density must equal $\Delta$ by  
conservation of
matter.  In
 the zero undercooling version, we know that if a steady state front  
forms it
must leave behind an aggregate density of some finite $\rho^o$. As  
$\Delta$ is
lowered below $\rho^o$ in the finite undercooling version, it will no  
longer be
able to form a travelling front, because its asymptotic
aggregate density is insufficient to support a travelling front in  
the zero
undercooling version.
Therefore a minimum undercooling $\Delta^*$ for travelling fronts in  
the finite
undercooling model is a signal of travelling fronts with an  
asymptotic walker
density $\rho^o$ in the zero undercooling version, and vice versa. We  
have
compared $\Delta^*$ and $\rho^o$ as measured in simulations  for  
various values
of  $\gamma$, and found a rough correspondence consistent with this  
argument
and the quality of our numerics.

Next we turn to the question of the existence of a Mullins-Sekerka  
instability.
The existence of a lower threshold for propagation $\Delta^*$  
indicates that,
in some sense, the front has a ``diffusion problem'' at low  
undercooling:
  it has difficulty attracting enough walkers to grow.
In the regime just above $\Delta^*$, we should not be surprised that  
the system
exhibits a Mullins-Sekerka instability,  as an outward perturbation  
can grow by
shielding neighboring regions.
The region below $\Delta^*$ in which the system cannot form a  
propagating front
is a region in which there must certainly be a Mullins-Sekerka  
instability
(compare with the instability of a motionless interface
\cite{MullinsSekerka,PhaseField3}). This instability persists in the  
moving
front as well, at least up until some velocity fast enough to damp  
it; a
phenomenon well known in the standard models of crystal growth.
In a model with $\Delta^* = 0$, there is no diffusion problem, and no  
effective
shielding can take place. We found stability in this case.

\section{Conclusion}
\label{lastSec}
The diffusion limited growth problem is a complex and widely studied  
exemplar
for pattern formation far from equilibrium.
The patterns formed can be quite regular, such as the organized  
crystals of the
Dendritic morphology, or  surprisingly complex, such as the fractal  
structures
of isotropic single walker DLA.

To understand the stochastic behavior of DLA like patterns,  
statistical
methods have to be used. For other microscopic growth models, like  
the Eden
model, ballistic deposition, molecular beam epitaxy etc., this  
statistical
approach in the form of nonlinear Langevin equations is quite  
successful in
determining the scaling properties of the pattern and the growth  
process.
Notable among them is the success of the noisy Burger's equation (KPZ  
equation)
for
studying kinetic roughening processes such as the Eden model.
While these models are all
local growth models, DLA is  non-local, which makes the
construction of a proper field theory for DLA much harder than
the local models.

Mean field models, as a first step towards a fully renormalized  
theory,
play a central role in the equilibrium theory of critical phenomena,  
and
the corresponding theory in nonequilibrium systems is becoming better
developed.
In spatio-temporal chaos, for example\cite{Gluckman}, in the
Kuramoto-Sivashinsky equation \cite{Hwa}, and in diffusion-limited  
reaction
systems \cite{ReactionDiffusion,ReactionDiffusion2} statistical  
averaging and
mean field modelling are being used to obtain nontrivial macroscopic  
behavior
from microscopic models.
Work on mean field models of diffusion limited growth
thus fits into a larger context of efforts to understand
pattern forming systems far from equilibrium,
and must be understood before steps going beyond mean field theory
can be taken.

Unlike most of the other cases, where the behavior of the mean field
theory (the deterministic part of Langevin field theory)  is
rather straightforward, the mean field theory of diffusion limited  
growth
is more involved. As we have shown before, a proper cutoff has
to be introduced to avoid singular behavior of the mean field theory
in low dimensions.  Our modified
threshold mean field theories can exhibit the morphology transition,
since they have the appropriate instability for the dendritic  
morphology.

In this paper, we have made a precise stability calculation for the  
threshold
mean field theory and for the first time established the existence of  
a
Mullins-Sekerka instability  for models with $\gamma >1$. This
instability puts the widely observed transition between the dendritic
morphology
(with concave envelope shape) and the dense branching
morphology (with convex envelope shape) on
firm theoretical ground. We have also demonstrated the absence of
such instability for the naive mean field theory without a cutoff.
The drastically different behaviors between the model with and  
without
a cutoff can be related to their
different behavior at zero undercooling and
  strongly reconfirms the existence of some sort of cutoff for
the correct mean field theory.

\acknowledgements
DR and HL were supported by NSF grant \#DMR 94-15460.


\begin{thebibliography}{10}

\bibitem{KKL}
D.~A. Kesser, J. Koplik, and H. Levine, Adv. Phys. {\bf 37},  255   
(1988).

\bibitem{solidification}
A. Dougherty and A. Gunawardana, Phys. Rev. E {\bf 50},  1349   
(1994).

\bibitem{electrodeposition}
Y. Sawada, A. Dougherty, and J.~P. Gollub, Phys. Rev. Lett. {\bf 56},   
1260
  (1986).

\bibitem{electrodeposition2}
D.~G. Grier, E. Ben-Jacob, R. Clarke, and L.~M. Sander, Phys. Rev.  
Lett. {\bf
  56},  1264  (1986).

\bibitem{bacteria}
E. Ben-Jacob, H. Shmueli, O. Shochet, and A. Tenenbaum, Physica A  
{\bf 187},
  378  (1992).

\bibitem{bacteria2}
E. Ben-Jacob, O. Avidan, A. Tenenbaum, and O. Shochet, Physica A {\bf  
202},  1
  (1994).

\bibitem{HS}
H. Hentschel and A. Fine, Phys. Rev. Lett. {\bf 73},  3592  (1994).

\bibitem{Cross}
M.~C. Cross and P.~C. Hohenberg, Rev. Mod. Phys. {\bf 65},  851   
(1993).

\bibitem{ShochetThesis}
O. Shochet, M.sc. thesis, Tel-Aviv University, 1992.

\bibitem{ShochetMorph}
O. Shochet {\it et~al.}, Physica A {\bf 181},  136  (1992).

\bibitem{ShochetMorph2}
O. Shochet {\it et~al.}, Physica A {\bf 187},  87  (1992).

\bibitem{coexist}
O. Shochet and E. Ben-Jacob, Phys. Rev. E {\bf 48},  R4168  (1993).

\bibitem{WittenSander}
T. Witten and L.~M. Sander, Phys. Rev. Lett. {\bf 47},  1400  (1981).

\bibitem{BLT}
E. Brener, H. Levine, and Y. Tu., Phys. Rev. Lett. {\bf 66},  1978   
(1991).

\bibitem{BLT2}
Y. Tu and H. Levine, Phys. Rev. A15 {\bf 45},  1044  (1992).

\bibitem{BLT3}
Y. Tu and H. Levine, Phys. Rev. A15 {\bf 45},  1053  (1992).

\bibitem{Tu}
Y. Tu, H. Levine, and D. Ridgway, Phys. Rev. Lett. {\bf 71},  3838   
(1993).

\bibitem{PhaseField}
J.~B. Collins and H. Levine, Phys. Rev. B {\bf 31},  6119  (1985).

\bibitem{PhaseField2}
J.~S. Langer,  in {\em Directions in Condensed Matter Physics},  
edited by
  Grinstein and Mazenko (World Scientific, 1986), pp.\ 164--186.

\bibitem{PhaseField3}
R.~J. Braun, G.~B. McFadden, and S.~R. Coriell, Phys. Rev. E {\bf  
49},  4336
  (1994).

\bibitem{Uwaha}
M. Uwaha and Y. Saito, Phys. Rev. A {\bf 40},  4716  (1989).

\bibitem{Keblinski}
P. Keblinski, A. Maritan, F. Toigo, and J. Banavar, Phys. Rev. E {\bf  
49},
  R4795  (1994).

\bibitem{LiuGoldenfeld}
F. Liu and N. Goldenfeld, Phys. Rev. A15 {\bf 42},  895  (1990).

\bibitem{TL}
Y. Tu and H. Levine, Phys. Rev. E {\bf 48},  R4207  (1994).

\bibitem{Graham}
R. Graham and A. Schenzle, Phys. Rev. A {\bf 26},  1676  (1982).

\bibitem{Shapiro}
E. Shapiro, Phys. Rev. E {\bf 48},  109  (1993).

\bibitem{Wulff}
G. Wulff, Z. Krist. {\bf 34},  449  (1901).

\bibitem{Dobrushin}
R. Dobrushin, R. Kotecky, and S. Shlosman, {\em Wulff Construction: A  
Global
  Shape from Local Interaction} (American Mathematical Society,  
Providence,
  Rhode Island, 1992).

\bibitem{Wolf}
D.~E. Wolf, J. Phys. A {\bf 20},  1251  (1987).

\bibitem{Wettlaufer}
J.~S. Wettlaufer, M. Jackson, and M. Elbaum, J. Phys A: Math Gen.  
{\bf 27},
  5957  (1994).

\bibitem{MarginalStability}
E. Ben-Jacobs {\it et~al.}, Physica D {\bf 14},  348  (1985).

\bibitem{pushPull}
G. Paquette, L. Chen, N. Goldenfeld, and Y. Oono, Phys. Rev. Lett.  
{\bf 72},
  76  (1994).

\bibitem{MullinsSekerka}
W.~W. Mullins and R.~K. Sekerka, J. Appl. Phys. {\bf 35},  444   
(1964).

\bibitem{Gluckman}
B.~J. Gluckman, C.~B. Arnold, and J.~P. Gollub, Phys. Rev. E {\bf  
51},  1128
  (1995).

\bibitem{Hwa}
C.~C. Chow and T. Hwa, Defect-Mediated stability: an effective  
hydrodynamic
  theory of spatiotemporal chaos, on xxx.lanl.gov as  
cond-mat/9412041, 1994.

\bibitem{ReactionDiffusion}
D. ben Avraham, M.~A. Burschka, and C.~R. Doering, J. Statistical  
Physics {\bf
  60},  695  (1990).

\bibitem{ReactionDiffusion2}
B.~P. Lee and J. Cardy, Phys. Rev. E {\bf 50},  R3287  (1994).

\end{thebibliography}

\end{document}